\documentclass[3p, 11pt]{elsarticle}
\usepackage{graphicx,amscd,amsmath,amssymb,verbatim}
\usepackage{amsfonts,epsfig}
\usepackage{mathptmx}
\usepackage{setspace}
\usepackage{epsfig}
\usepackage{url}
\usepackage{multirow}
\usepackage{color}
\usepackage{subfigure}

\begin{document}

\journal{Elsevier}

\begin{frontmatter}

\title{Spatial patterns of random walkers under evolution of the 
attractiveness: persistent nodes, degree distribution, and spectral properties}

\author{Roberto da Silva} 

\address{Institute of Physics, Federal University of Rio Grande do Sul,
Av. Bento Gon\c{c}alves, 9500, Porto Alegre, 91501-970, RS, Brazil
{\normalsize{E-mail:rdasilva@if.ufrgs.br}}}


\begin{abstract}

In this paper we explore the features of a graph generated by random walkers
with nodes that have evolutionary attractiveness and Boltzmann-like
transition probabilities that depend both on the euclidean distance between
the nodes and on the ratio ($\beta $) of the attractiveness between them. We
show that persistent nodes, i.e., nodes that never been reached by random
walker in asymptotic times are possible in the stationary case differently
from the case where the attractiveness is fixed and equal to one for all
nodes ($\beta =1$). Simultaneously, we also investigate the spectral
properties and statistics related to the attractiveness and degree
distribution of the evolutionary network. Finally, we study a crossover
between persistent phase and no persistent phase and we also show the
existence of a special type of transition probability that leads to a power
law behaviour for the time evolution of the persistence.
\end{abstract}

\end{frontmatter}

\tableofcontents

\setlength{\baselineskip}{0.7cm}

\section{Introduction}

\label{Sec:Introduction}

The idea of preferential attachment of Barabasi and Albert \cite%
{Barabasi1999,Albert2002} has brought a revolution to the study of complex
systems mainly by the simplicity of the idea and its wide applicability
which goes from web, epidemics, metabolic networks, scientific
collaborations, human mobility and so on. Thus, a really wide scope of the
extensions of this and many other ideas in network science have been
developed \footnote{%
http://barabasi.com/networksciencebook/}.

An interesting point in the theory of evolutionary networks is related to
how random walks with properties on the edges or on the nodes can perform
this evolution starting from initial condition where only nodes exist and
are distributed in a two-dimensional surface.

In this context certain known properties of random walks as first time
passage \cite{Feller1968} and persistence \cite%
{Majundar1996,Derrida,RdasilvaPersistence} should be closely linked with
properties of the generated graph formed by the path (sequence of the
visited edges and nodes) of the random walker taking into account for a
dependence on the spatial distribution of the nodes and their peculiarities.
Following this idea, the walker should be guided by the euclidean distance
between the nodes. Moreover these nodes also should have a kind
attractiveness characterized for example by the frequency which they are
visited along the time evolution, remembering in some sense of preferential
attachment.

In this context the internet is a good example. We can imagine that
individuals have some distance in relation to some topics. For example, one
has some preference list and navigates in the internet, reading news about
science, maybe politics. Thus we consider that she(he) keeps a short
distance in relation to these topics. On the other hand, she (he) keeps high
distances in relation to topics such as soccer (not me), religion, social
gossips, so that the access probability is initially small. However since
some topics have been accessed, they call attention even of non-usual users
increasing the access probabilities and contributing to the formation of the
so-called \textquotedblleft trending topics\textquotedblright .

Looking at the graph properties, we can highlight some basic properties as
the degree distribution, distribution of accesses but also spectral
properties as the density of eigenvalues of matrices related to adjacency
matrix of the random walk generated graph. The idea of considering a random
walk in a set of random points in a two-dimensional space brings a lot of
interesting discussion in Physics and an important question refers to the
problem of a particular node not being visited at large times, or
translating to a spin system the question change to what the probability
that a particular spin does not change its state until time $t$.

Such concept was deeply studied by many authors, considering dynamics at
temperature $T=0$, known as coarsening dynamics (see \cite{Derrida}). A
simple dynamics in this context, for example considers that if a flip of a
particular spin state interacting only with its first neighbours, decreases
the energy system, its state must be changed otherwise it changes with
probability $1/2$. In this case the fraction of spins with unaltered state
from the beginning to the time $t$ in one, two, or three-dimensional
lattices decays as power law as function of time

\begin{equation}
Pers(t)\sim t^{-\theta }\text{.}  \label{Eq.power_law_decay}
\end{equation}%
Such behaviour is exponentially stretched for $T\neq 0$. This concept is
known as local persistence. Similarly, this concept has a global version 
\cite{Majundar1996}. In this case, it can be shown that for systems at
critical temperature, the probability of magnetization (essentially the sum
of spin states divided by the system size) does not have changed its sign
until time $t$ also decays as a power law given in Eq. \ref%
{Eq.power_law_decay}.

So we are interested in persistent sites in the graph formed by the random
walk, or more precisely if we have a power law decay for the persistence
which means that some nodes are persistent at large times. Alternatively, if
the decay is exponential, in the process we have a fixed fraction of
persistent sites in the limit:

\begin{equation}
\lim_{t\rightarrow \infty }Pers(t)=p_{\infty }  \label{Eq:conditiion}
\end{equation}%
with $p_{\infty }>0$. Starting from this idea, let us imagine that a
navigation of a particular people in the web can be mimicked as a random
walk with some idiosyncrasies. In this random walk, in general, we suppose
that this person has some initial distances in relation to the topics,
however after a visit to a particular topic, the attractiveness of this
site, which was not even the most important, has changed given its visit,
and the probability of visit to this topic now, not only depends on the
distance but also depends of its attractiveness that initially was the same
for all sites.

So if we consider that walk is governed initially by the distances of the
walker in relation to nodes and after some time the incidence of the walker
on the nodes make them more attractive which works as a mechanism to change
the effects of the distance the transition probabilities among the different
pairs of nodes, what is the properties of the graph formed during the
evolution of this peculiar random walk? In a more abstract point of view, we
can imagine a mathematical modelling where a walker paints an edge (if this
does not exist) when transits between two nodes and the growing graph
obtained from this process (nodes + painted edges) can be studied.

In the literature some authors have explored similar models such for example
the step-by-step random walk network model described in \cite{Yang2008}, and
preferential network random walk studied in \cite{Mehraban2012}. All these
models have a similar dynamics but different aims. Differently from our
original idea to this current work, these models start from a fully
connected network of a certain number of nodes $m$. In \cite{Yang2008}, a
new node is created which will be connected to the certain number of nodes $%
n\leq m$. These nodes which will linked to the new created node are
determined by a step by step self-avoiding random walk defined on the
current network. Actually this model is a simple mechanism for generating
scale-free networks.

Alternatively, \cite{Mehraban2012} considers multiple random walkers over a
fully connected network. At time-step, each walker chooses one of its
neighbours with a certain probability and transits to the chosen neighbour.
When the walker passess by the link the transition probability is decreased
(destructive approach) or enlarged (constructive approach).

In this work, our idea is quite different. We explored the properties of a
random walk where the distance between nodes as well as their attractiveness
are taken into account and its connexion with the spectral properties of the
adjacency matrix of the generated network and other properties of this
network.

More precisely, we consider $N$ points randomly distributed in a unity
square. Starting from an initial point randomly chosen, a single random walk
can transit to other randomly chosen node. But differently from \cite%
{Mehraban2012} we consider that potential (or attractiveness) of the nodes
(not links as in \cite{Mehraban2012}) achieved by the random walker are
increased by the incidence of the walker and the transition probability
between two nodes follows a Boltzmann form which depends not only on the
distance between these nodes but also on the ratio of the potentials of
these two nodes.

In our analysis, we performed Monte Carlo (MC) simulations. First, we
performed an analysis where the attractiveness of the nodes remain fixed and
the same for all nodes. In this case only the distance is important and we
compared the existence of persistent nodes with the situation where the
attractiveness is incremented according to the incidence on the nodes.

Additionally, we study the spectral properties of a matrix that depends on
the adjacency matrix of formed graph obtained by our walker with this
special random walk. More precisely we obtained the density of eigenvalues
of this matrix comparing with expected law (a variation of semi-circle law)
valid for random graphs (Erd\H{o}s--R\'{e}nyi model) exactly for connexion
probability $p=1/2$ . Finally we also study the effects on the existence of
the persistent nodes, by considering some modifications of the transition
probability with Boltzmann and non-Boltzmann functional dependence.

In the next section (Sec. \ref{Sec:The_model_and_methods}) we present the
model, the numerical methods and some details of the random matrices theory
(RMT) to be employed in the analysis of the results of this work.
Additionally, we describe a simple mean-field method to describe the
persistence probability for a comparison with numerical results.

In section \ref{Sec:Simulational_results} we present our results obtained
via numerical simulations. Finally in Sec. \ref{Sec:Conclusions} we present
our summaries and the main conclusions.

\section{The model and methods}

\label{Sec:The_model_and_methods}

In the next subsection we present the model to be used to describe the
spatial exploration with our random walk based on the distance and
attractiveness ratio among the pairs of nodes. Following, we also explore
some details of analysis and numerical methods to be used in the analysis of
the problem and finally in the last subsection, a simple mean field result
is obtained when the attractiveness is constant and the same for all nodes
in order to compare with numerical results for the existence of the
persistent nodes.

\subsection{ The model}

\label{Sec:The model}

Let us consider a graph generated by the time evolution of a random walk
defined on the nodes $(x_{i},y_{i})$, with $i=1,2,...N$, randomly
distributed in the two-dimensional unity square $[0,1]^{2}$. There are no
edges initially.

In this scenario, we add a special ingredient, where all nodes start with
initial potential $\varphi _{i}=1$, which means (by thinking in the context
of social physics, or web applications and so on) its attractiveness. In
each run we choose a node $j=$ $1,...,N$, where our walker starts. Another
node $k=1,...,N$ is randomly chosen and the walk jumps to the node $k$ with
probability which follows a Boltzmann weight:%
\begin{equation}
p_{jk}=p(j\rightarrow k)=e^{-N^{\delta }\beta _{jk}d_{jk}}
\label{Eq:transition}
\end{equation}%
where $d_{jk}=d_{kj}=\sqrt{(x_{j}-x_{k})^{2}+(y_{j}-y_{k})^{2}}$ , with $%
0\leq \delta $. This indicates that probability depends on Euclidean
distance between the nodes. A walker, for which the jump from one site to
another is performed with probability proportional to $\exp (-E(d)/T)$ where 
$E(d)$ is an arbitrary cost that depends on the hop distance $d$ was studied
in \cite{MartinezI}. These authors show to exist a glass transition for a
critical value of $T$. In other work \cite{MartinezII}, considering
deterministic random walks in one-dimensional environment (the walker goes
to the nearest site) with a memory $\mu $, the authors show a crossover
between localized and extended regimes at the critical memory $\mu _{c}=\log
_{2}N$.

Thus, we propose to consider a kind of \textquotedblleft local network
temperature\textquotedblright , considering the thermodynamic motivation of
the problem: 
\begin{equation}
\beta _{jk}=\frac{\varphi _{j}}{\varphi _{k}}  \label{Eq:beta}
\end{equation}%
i.e., the higher is the attractiveness of the arrival node is in relation to
starting node, the smaller is the $\beta $, and therefore larger is the
probability to jump between the nodes.

So by finally capturing the idea of preferential attachment, nodes reached
by the walk become more probable to be reached, since every time that a node
is crossed by random walk we increment the potential in one unit: $\varphi
_{new}=\varphi _{old}+1$. This modelling should be extended for example in
the modelling of knowledge acquisition dynamics by professionals in a work
network, since in general, a scientist for example, has a set of topics
which keeps a certain distances from them, as we evolve in the studied topic
and the nodes are visited with some frequency, in general, users tend to
visit the same topics that other people (or even them) have already been
accessed.

Thus in this work, we propose such model which combines two important ideas:
the affinity that people have with topics in a network, for example defined
by the euclidean distances $d(i,j)$ which is compensated by the individual
popularity of node (its attractiveness) which is modelled by the number of
accesses.

Finally, in Eq. \ref{Eq:transition}, we consider a scaling factor $N^{\delta
}$ which controls the importance of distance and attractiveness effects in
the argument of the Boltzmann weight. If for example $\delta =0$, we have,
initially since $\varphi _{i}(t=0)=1$, for all sites, the inequality: $e^{-%
\sqrt{2}}<p(j\rightarrow k)<1$. Since $e^{-\sqrt{2}}\approx 0.24$, we have
large probabilities to jump which is very different from $\delta =1/2$. In
this case, for $N=100$ sites, we have $p_{\min }=$ $e^{-10\sqrt{2}}\approx
7.\,\allowbreak 2\times 10^{-7}$.

\subsection{Monte Carlo Simulations and some numerical analysis}

\label{Sec:Monte_Carlo_Simulations}

In this paper, we perform numerical MC simulations to study the statistical
fluctuations related to properties of our preferential random walk and
consequently of generated graph. An initial set of points is random and
uniformly sorted in the two-dimensional unit square. So we choose an initial
random node, and the graph is built as the nodes (and edges) are travelled
by the random walk.

Here is appropriated to consider simulation divided by turns, i.e., the
system evolves and for each jump we paint the edge which governs our network
evolution. After $N$ jumps (or simply attempts) one turn is completed in a
numerical simulation for example. So one time unit, $t=1,2,3...,t_{\max }$ ($%
t$ turns) supposes that $\tau =Nt$ jumps were executed by the walker. For
each turn you can imagine that one other walker starts a new walk
considering that the previous walker has evolved the network built up to
that moment according to your path.

In order to compute averages, we repeat the $N_{run}$ different evolutions,
and the averages of the different amounts were obtained over different
sources of variations: different set of $N$ points distributed in the unit
square, different initial selected point, and different random numbers used
during the evolution to perform the transitions among the nodes.

Our model, which adds two important concepts: spatial exploration and
network formation by a random walk, suggests important questions about the
effects of the attractiveness of nodes on the network properties generated
and about statistics of accesses to these nodes. Among the very interesting
statistics we can ask about the probability a node has of having not been
reached up to time $t$, which is quantified by the previously described
amount, known by statistical physicists as local persistence, which from
this computational scenario is here calculated by 
\begin{equation}
p(t)=\frac{1}{N}\frac{\sum\limits_{j=1}^{N_{run}}n_{j}(t)\text{ }}{N_{run}}
\label{Eq:persitence}
\end{equation}%
where $n_{j}(t)$ is the number of unreachable nodes at time $t$, at in the $%
j $-th run. Intuitively, obtaining such amount it must bring information
about the evolution and the final stage of the emergent network.

Simultaneously, in this paper we also analyze the spectral properties of
generated graph by random walker. So, for a better numerical results, we
propose to consider the information from the more appropriate matrix that
describes our graph, which is defined by:%
\begin{equation*}
C=\frac{1}{N}A^{T}A
\end{equation*}%
were $A$ is a simple mapping from adjacency matrix $M$, where $M_{ij}=1$ if
node $i$ is linked to node $j$ and 0 otherwise, defined by $%
A_{ij}=2M_{ij}-1=\pm 1$, and in this particular case $C=\frac{1}{N}A^{2}$.
If $A_{ij}$ is a random variable with $A_{ij}=0$ and $\left\langle
A_{ij}^{2}\right\rangle -\left\langle A_{ij}\right\rangle ^{2}=1$, we will
enunciate which is a known extension of the famous semi-circle law.
Considering the cumulative distribution of eigenvalues of $C$,

\begin{equation*}
F_{N}(\lambda )=\frac{1}{N}\#\{\lambda _{i},\lambda _{i}<\lambda \}
\end{equation*}%
where $\#\{\lambda _{i},\lambda _{i}<\lambda \}$ denotes the number of
eigenvalues which are smaller than $\lambda $, the density of eigenvalues
defined by:

\begin{equation*}
\rho _{C}(\lambda )=\lim_{N\rightarrow \infty }\frac{dF_{N}(\lambda )}{%
d\lambda }
\end{equation*}%
is given by \cite{Sengupta1999} 
\begin{equation}
\rho _{C}(\lambda )=\left\{ 
\begin{array}{lll}
\frac{\sqrt{(\lambda _{+}-\lambda )(\lambda -\lambda _{-})}}{2\pi \lambda }
&  & \lambda _{-}\leq \lambda \leq \lambda _{+} \\ 
&  &  \\ 
0 &  & \lambda <\lambda _{-}\text{,\ }\lambda >\lambda _{+}\text{ }%
\end{array}%
\right.  \label{Eq:modified_semi_circle_law}
\end{equation}%
where $\lambda _{-}=0$ and $\lambda _{+}=4$.

So by using the Jacobi%
\'{}%
s method to diagonalize $C$ \footnote{%
Using $C$ and not $A$ directly is more interesting since according to method
employed to determine the eigenvalues, $A$ seems to be more sensitive than $%
C $ and some numerical problems were detected in the first case.}, we
compute the eigenvalues as the graph generated by the random walker evolves.
In order to compute the numerical $\rho _{C}(\lambda ;t)$ we accumulate for
each instant $t$ of MC simulation, $N_{run}$ sets of $N$ eigenvalues.

The idea is to observe the similarity and distortions between the real
eigenvalues distribution and theoretical prediction. This last supposes that
graph is a genuine Erdos-Renny graph \cite{Erdos1959} with parameters $N$
and $p$, i.e., there is a edge between any pair of nodes of the network with
probability $p$ and a total of $N$ nodes. We believe that our graph must
pass by this situation (where the nodes have approximately the same
coordination in average) for intermediate times of the evolution. But in
general we must observe the eigenvalues that scape from the law \ref%
{Eq:modified_semi_circle_law}, but how the eigenvalues scape from this
theoretical bulk, is deeply related to walker cover the nodes and compose
the graph. \ 

\subsection{Mean-field regime}

\label{Sec:Mean-Field-regime}

Let us to explore the process in a first special kind of spacial exploration
of our walker where $\beta $ is constant. Particularly when the
attractiveness is constant and equal for all nodes during the time evolution
corresponds to $\beta =1$. Particularly, we consider that in a special
regime where we exchange $d_{ij}$ by $\left\langle d\right\rangle $. At time 
$t>0$, the probability of any site to be reached by the walk, in a kind of
\textquotedblleft mean field regime\textquotedblright\ is given by 
\begin{equation*}
p=\frac{e^{-\beta N^{\delta }\left\langle d\right\rangle }}{N}
\end{equation*}%
since we randomly choose a site we jump to it with probability $e^{-\beta
N^{\delta }\left\langle d\right\rangle }$, where 
\begin{equation*}
\begin{array}{lll}
\left\langle d\right\rangle & =\lim_{N\rightarrow \infty }\frac{1}{N(N-1)/2}%
\sum_{i<j}d_{ij} &  \\ 
&  &  \\ 
= & \int\limits_{0}^{1}\int\limits_{0}^{1}\int\limits_{0}^{1}\int%
\limits_{0}^{1}dx_{i}\ dx_{j}\ dy_{i}\ dy_{j}\left[
(x_{i}-x_{j})^{2}+(y_{i}-y_{j})^{2}\right] ^{1/2} &  \\ 
&  &  \\ 
\approx & 0.519 & 
\end{array}%
\end{equation*}

If our time is computed in turns of $N$ jump trials, the probability of a
site is not reached by the walk up to time $\tau =tN$, is%
\begin{equation*}
\Pr (\tau =tN)=\hat{p}=(1-p)^{tN}
\end{equation*}%
and the probability that $N_{pers}$ sites are persistent at time is%
\begin{equation*}
\Pr (N_{pers})=\frac{N!}{N_{pers}!(N-N_{pers})!}\hat{p}^{N_{pers}}(1-\hat{p}%
)^{N-N_{pers}}
\end{equation*}%
and the average number of persistence sites at time $\tau =tN$, is $%
N_{pers}=N\hat{p}$ and a good approximation to the probability of a site is
persistent at time $\tau =tN$ is the fraction 
\begin{equation*}
\begin{array}{lll}
Pers(t) & = & \frac{\overline{N_{pers}}}{N}=\hat{p} \\ 
&  &  \\ 
& \approx & \left( 1-\frac{e^{-0.519N^{\delta }\beta }}{N}\right) ^{Nt}%
\end{array}%
\end{equation*}

For fixed $N$ we have an exponential decay for $Pers(t)$ and the sites must
be explored after some time and we do not expect persistent sites for $%
t\rightarrow \infty $. For example, particularly for $\delta =0$, we have 
\begin{equation*}
Pers(t)\sim \exp (-e^{-0.519\beta }t)\text{.}
\end{equation*}%
\ 

This simple mean field analysis suggests that no persistent sites are
expected at large times. The important question if this mean field regime
corresponds to similar behaviour to be found in the case where $\beta $
evolves according to evolution of potentials. The answer is no! And our
numerical simulations will corroborate such study by showing that we expect
persistent sites when the nodes have different attractiveness which changes
over time by the incidence.

We also complete this study in the next section looking the effects of $%
\delta $ on the persistent sites as well as the evolution of formed graphs
by the walker. Moreover we study the spectral properties of this graphs
under light of random matrix theory and finally some results about other
transition probabilities are presented.

\section{Simulation Results  }

\label{Sec:Simulational_results}

In this section we present our numerical results.  

\subsection{Persistent nodes: degree and attractiveness distributions}

In this paper we performed MC simulations considering initial set of points
(nodes) randomly distributed in the two-dimensional unit square. So we
choose an initial random node, and the graph is built as the nodes (and
edges) are travelled by the random walk and we perform $t_{\max }$ turns
(each turn composed by $N$ jumps or attempts). This process are repeated $%
N_{run}$ times to average the quantities obtained in this paper.

As suggested by preliminary results obtained in mean-field regime, if we
keep $\beta $ constant ($\varphi _{i}$ constant and the same for all nodes)
the random walker explores the space and no persistent site is expected at
large times. Sure the exploration time (necessary time for all sites to be
visited) depends on $\delta $. By starting our simulations we elaborated a
plot that computes the persistence considering $N_{run}=400$ runs.

We compute the persistence $p(t_{\max })$ as function of $\delta $, in order
to qualitatively check the mean-field results. Here $t_{\max }$ is the last
turn of evolution.

\begin{figure}[th]
\begin{center}
\includegraphics[width=%
\columnwidth]{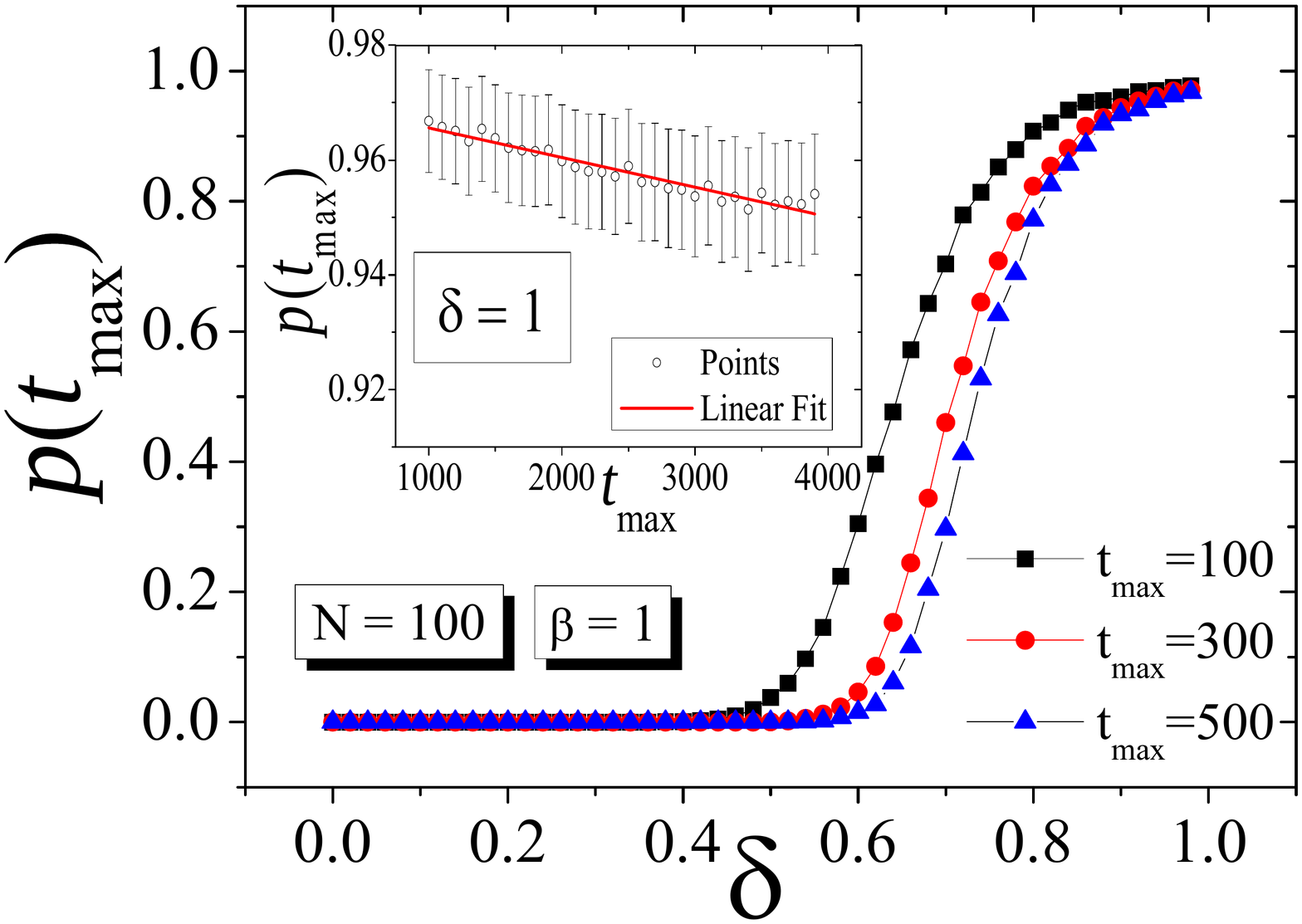}
\end{center}
\caption{Persistence $p(t_{\max })$ as function of $\protect\delta $ for
different values of $t_{\max }$. We can see that after a $\protect\delta %
_{c}(t_{\max })$, $\ $we have $p(t_{\max })>0$. It is interesting to note
that $p(t_{\max })$ is larger as $t_{\max }$ increases. For example, in the
worst case, $\protect\delta =1$, for $t_{\max }=1000$, $p(t_{\max })\approx
1 $. According to mean-field prescription, $p(t_{\max })$ must decrease as
function of $t_{\max }$ as can be observed in the inset plot however the
decay is very slow and we estimate $p(t_{\max })=0$ for $\protect\tau %
\approx tN\approx 10^{8}$ steps of the walk since the slope is $O(10^{-6}$).
For these simulations we used $\protect\beta =1$ and $N=100$. For estimating 
$p(t_{\max })$ we used $N_{run}=400$ different runs for a good estimate. }
\label{Fig:Corroborating_exponencial_decay_MC}
\end{figure}

In Fig. \ref{Fig:Corroborating_exponencial_decay_MC} we plot the value of
persistence $p(t_{\max })$, estimated as function of $\delta $ for different 
$t_{\max }$. We can see that after a $\delta _{c}(t_{\max })$, $\ $we have $%
p(t_{\max })>0$. We observe that $p(t_{\max })$ point is larger as $t_{\max
} $ increases. For example, in the worst case, $\delta =1$, for $t_{\max
}=1000 $, $p(t_{\max })\approx 1$. But according to mean-field prescription, 
$p(t_{\max })$ must decreases as function of $t_{\max }$ as can be observed
in the inset plot however the decay is very slow and we expected $p(t_{\max
})=0$ for $\tau \approx tN\approx 10^{8}$ steps of the walk since the slope
is $O(10^{-6}$). For these simulations, we fix $\beta =1$ and $N=100$. For
estimates of $p(t_{\max })$, we used $N_{run}=400$ runs.

\begin{figure}[th]
\begin{center}
\includegraphics[width=1.0%
\columnwidth]{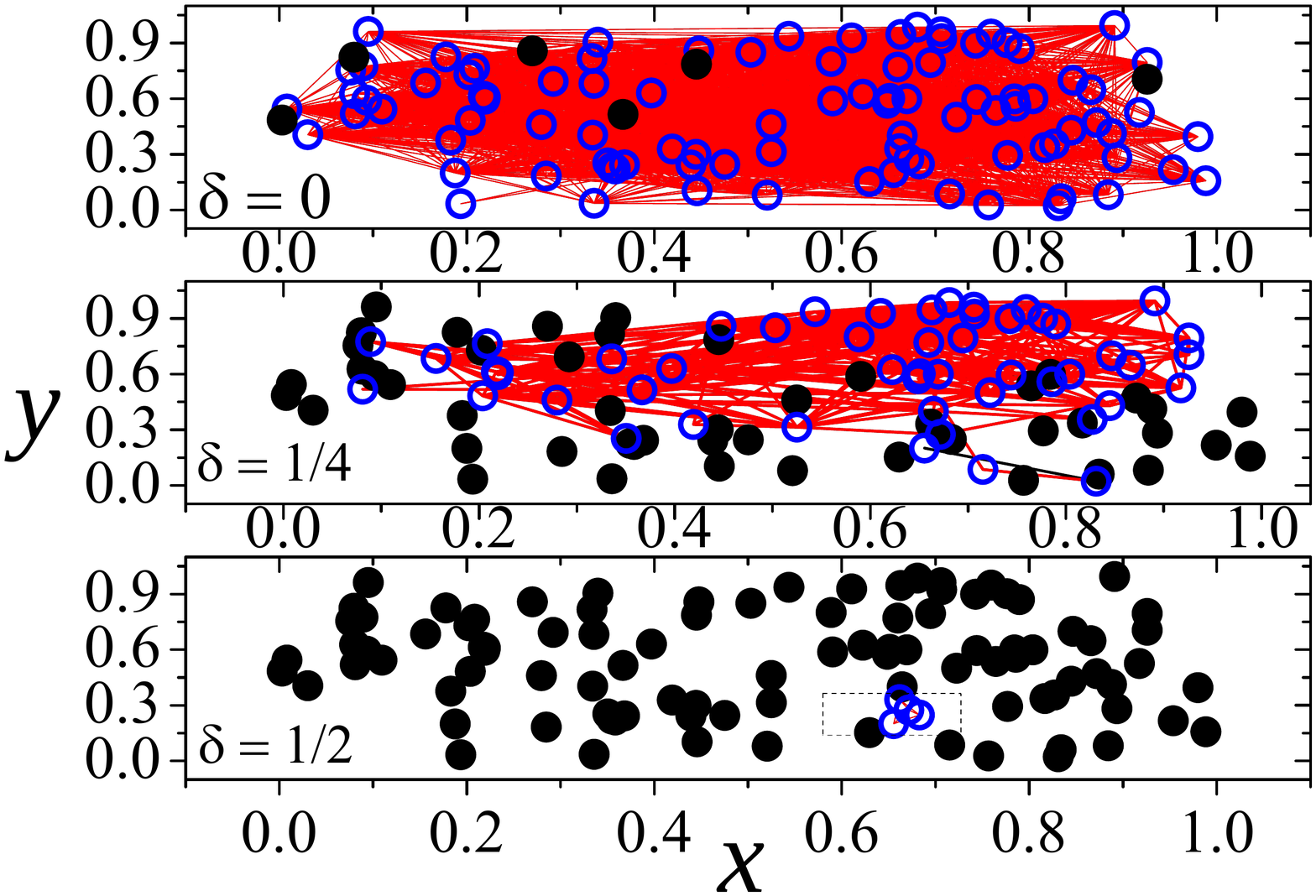}
\end{center}
\caption{Generated graphs for $t_{\max }=1000$ turns considering
evolutionary $\protect\beta $ for different values of $\protect\delta $.
Full circles correspond to persistent sites while the open circles
correspond to sites which already were visited. }
\label{Fig:graphs_different_deltas}
\end{figure}

So the question how the evolutionary attractiveness of the nodes affects the
graph evolution. Now we consider $\beta $ given in Eq. \ref{Eq:beta}. We
start our analysis looking to the graph at large times ($t_{\max }=1000$
steps). As can be observed in Fig. \ref{Fig:graphs_different_deltas} we show
the graphs obtained for a specific run and in all situations we have
persistent sites even for $\delta =0$. This suggests that persistence has no
exponential decay to zero at large times and we must expect an asymptotic
convergence to value $p_{\infty }>0$.

It is important to call the attention that for $\delta =1/2$, the walker is
confined in no more than 3 sites. So let us now concentrate our study in the
worst case for the study of the persistence, $\delta =0$. So we elaborate a
plot of the time evolution of $p(t)$, for different size systems (number of
nodes), for $\delta =0$. Here $p(t)$ was estimated by using $N_{run}=1000$
runs.

\begin{figure}[th]
\begin{center}
\includegraphics[width=1.0%
\columnwidth]{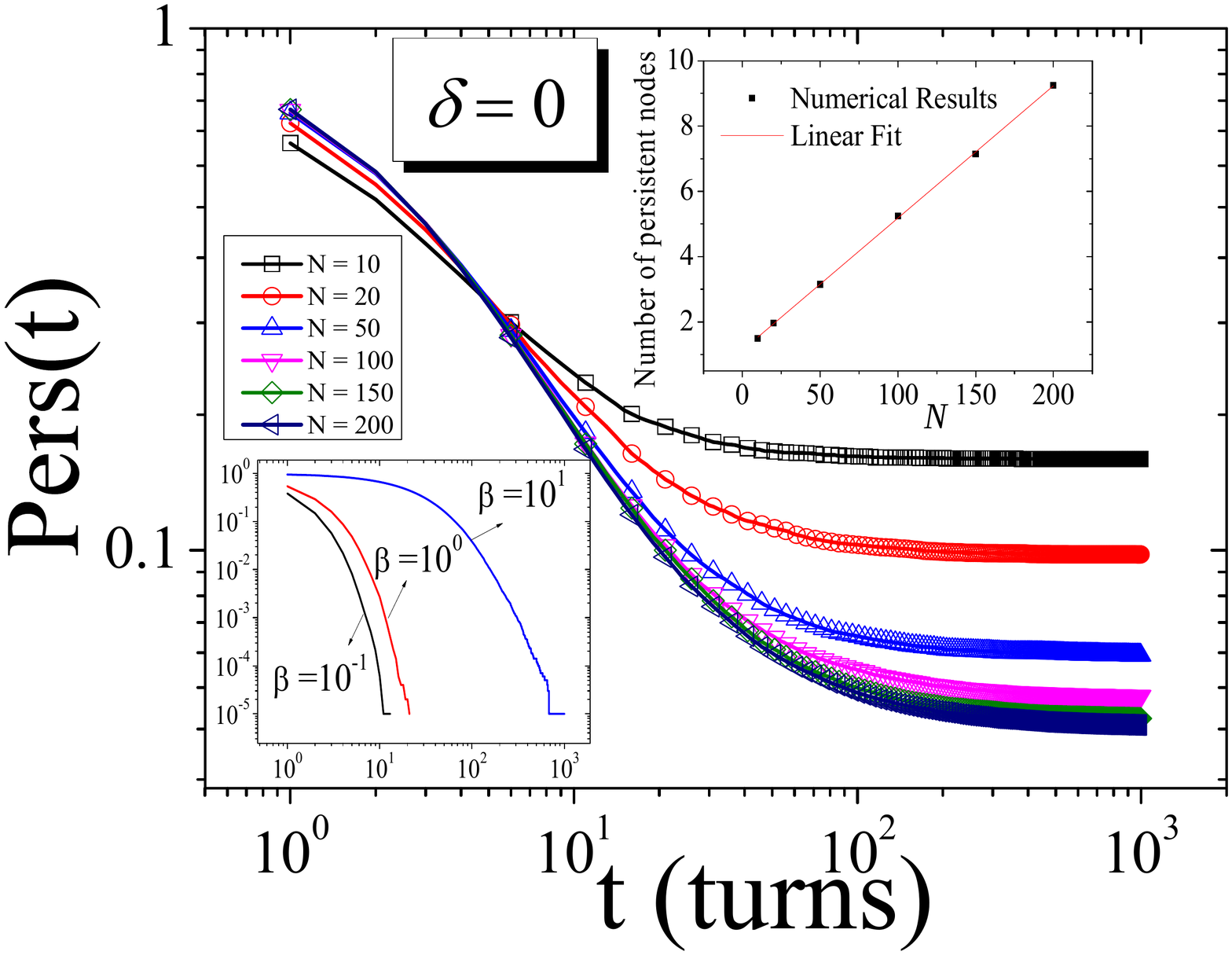}
\end{center}
\caption{Time evolution of the persistence, for $\protect\delta =0$, for
different size systems in log-log scale. We can observe that for large times
the persistence asymptotically goes to to a constant $p_{\infty \text{ }}>0$%
. The upper inset plot shows the number of asymptotic persistent sites ($%
n_{\infty }=Np_{\infty \text{ }}$) as function of $N$. We note a linear
dependence (linear fit corresponds to red line) with a slope $\approx 0.04$.
Just for a comparison, the time evolution of persistence is also shown for $%
\ $some different fixed $\protect\beta $ valu es (lower inset plot) which
corroborates the exponential explorations of space, as predicted in
mean-field regime. }
\label{Fig:time_evolution_persistence}
\end{figure}

Time evolution of the persistence, for $\delta =0$, for different size
systems in log-log scale is shown in Fig. \ref%
{Fig:time_evolution_persistence}. We can observe for large times the
persistence asymptotically goes to a constant $p_{\infty \text{ }}>0$. The
upper inset plot shows the number of asymptotic persistent sites ($n_{\infty
}=Np_{\infty \text{ }}$) as function of $N$. We observe a linear dependence
(linear fit: continuous curve in red) with a slope $\approx 0.04$. Just for
a comparison, the time evolution of persistence is also shown for different
values for $\beta $ fixed during the simulation (lower inset plot). This
corroborates the previous exponential prediction in mean-field regime. So it
is interesting that even for $\delta =0$ we have an asymptotic number of
sites that never are passed through by the walker as suggested by Fig. \ref%
{Fig:graphs_different_deltas}.

The model is also suggesting an important point, the distance can delay the
space exploration but cannot confine the random walk. It only happens if the
attractiveness of the nodes evolves according to the dynamics here proposed.
Thus, it is interesting to better investigate the statistics related to
distribution of attractiveness and degree distributions of the nodes in the
network. Let us start by the degree distribution. Defining $f(k)$ the
frequency of nodes with degree equal to $k$, we look at distribution for
different turns. For that we separate our simulations again in a) $\beta =1$
and b) $\beta $ evolutes according to increment of the potentials, since we
want to check the differences and similarities of $f(k)$ in both situations.

\begin{figure}[th]
\begin{center}
\includegraphics[width=1.0\columnwidth]{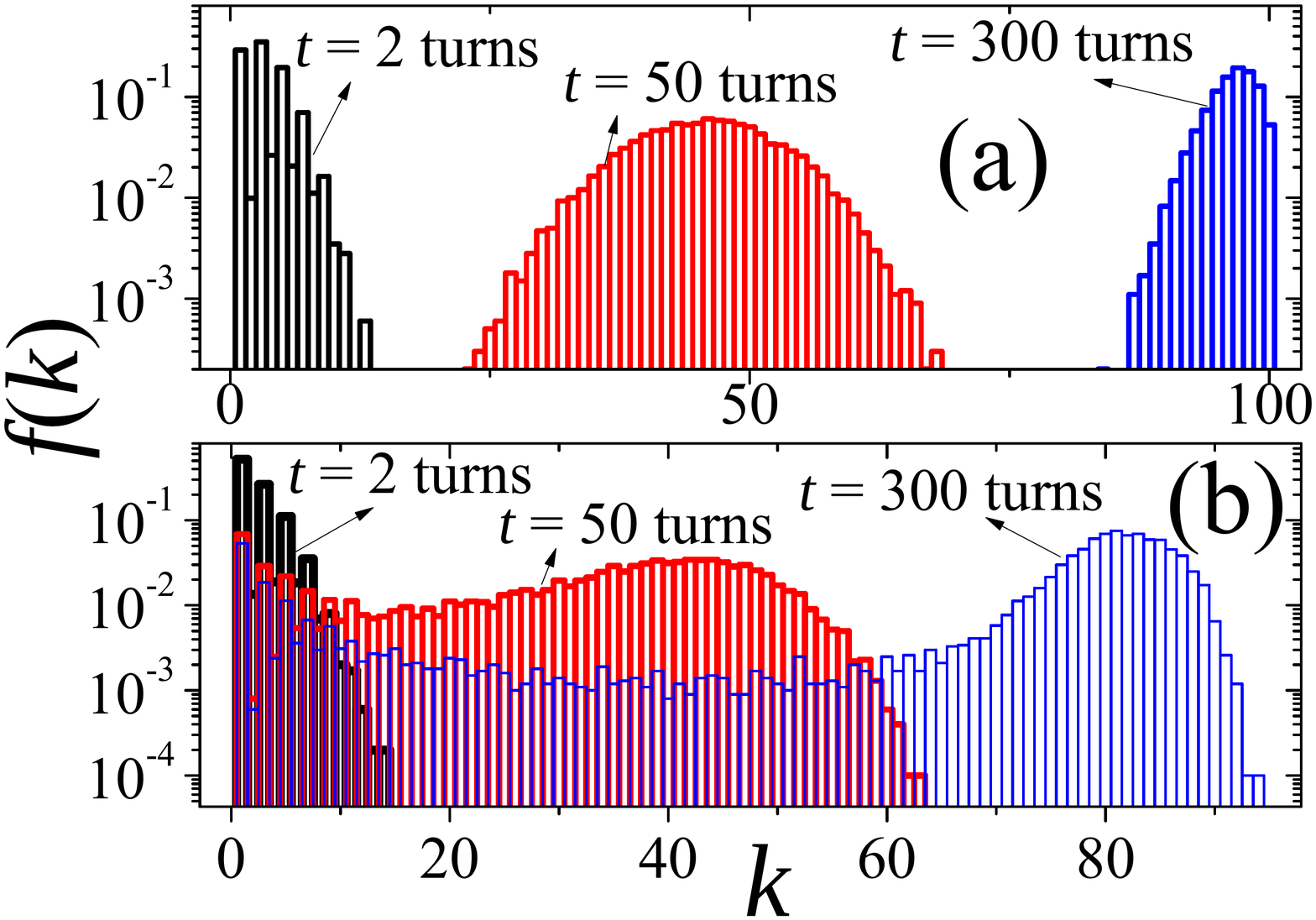}
\end{center}
\caption{Degree distribution for different times. (a) $\protect\beta =1$
(attractiveness fixed and equal to all nodes) (b) $\protect\beta $ evolves
over time according to the attractiveness evolution. }
\label{Fig:degree_distribution}
\end{figure}

In Fig. \ref{Fig:degree_distribution} we show the degree distribution for
three different instants: $t=2$, $50$, and $300$ turns. In (a) we can
observe (in mono-log scale) that for $t=50$ we have a gaussian distribution
of nodes, which suggests a formation of a random graph (Erd\H{o}s--R\'{e}nyi
model) with $p=1/2$, i.e., the probability that two randomly chosen nodes
are connected is equal to 1/2.

We can observe that this stage is not observed in the evolution of degree
distribution related to evolution of attractiveness (plot (b) in Fig. \ref%
{Fig:degree_distribution}). We used $N_{run}=1000$ runs and $N=100$, which
means that we have $1000$ sets of $100$ values of degrees to elaborate these
histograms. Actually, in the beginning of evolution, in both cases the
degrees are similarly distributed with low values, since theoretically $%
f(k,t=0)=\delta (k-0)$. As the walker evolves, what occurs is different: In
(a) we transit to a random graph with $p=1/2$ up to converge (at large
times) to a complete graph (all nodes are connected to all other nodes)
since $f(k,t\rightarrow \infty )=\delta (k-N)$.

Differently in (b) the system does not transit to the random graph with $%
p=1/2$ in intermediate times and does not converge to a complete graph,
since a selection of nodes as the system evolves over time, generates an
interesting shapes for distribution where nodes with high degree and low
degree are highly frequent with interesting valley for the intermediate
degrees of the nodes.

\begin{figure}[th]
\begin{center}
\includegraphics[width=1.0\columnwidth]{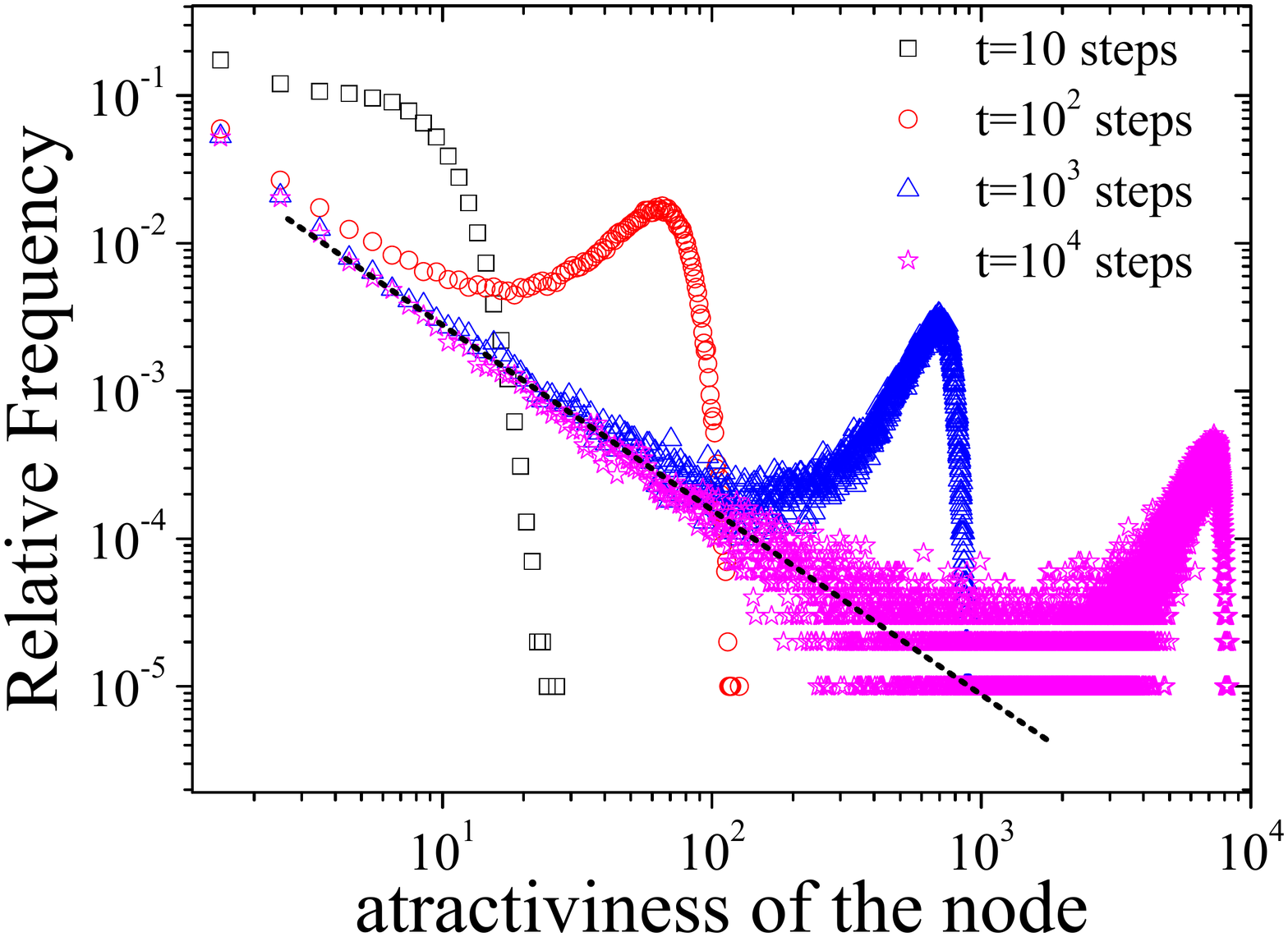}
\end{center}
\caption{Attractiveness distribution for different times in log-log scale.
The distribution is governed by a power law distribution but as time
increases, the increment of attractiveness leads to a strong deviation of
this power law since some of them become \textquotedblleft trading
topics\textquotedblleft\ in the generated network.}
\label{Fig:Atractiviness_distribution}
\end{figure}

The reason of this peculiar distribution found in (b) can be corroborated
looking for the attractiveness distribution. The distribution is governed by
a power law distribution as the system evolves, however the increment of
attractiveness leads to a strong deviation from this power law for nodes
with high attractiveness as can be observed in Fig. \ref%
{Fig:Atractiviness_distribution}.

\subsection{Spectral analysis \ }

We observed that when the attractiveness evolves over time we do not observe
a transition from $f(k,t=0)=\delta (k-0)$ to $f(k,t\rightarrow \infty
)=\delta (k-N)$ passing through an intermediate stage where the degrees of
the nodes follow a Gaussian distribution. We consider that a
non-conventional but an interesting and alternative way to describe the
features of a graph originated by the random walk is to analyze the spectral
properties of a matrix $C$ built from adjacency matrix as discussed in
section \ref{Sec:Monte_Carlo_Simulations}.

\begin{figure}[th]
\begin{center}
\includegraphics[width=1.0\columnwidth]{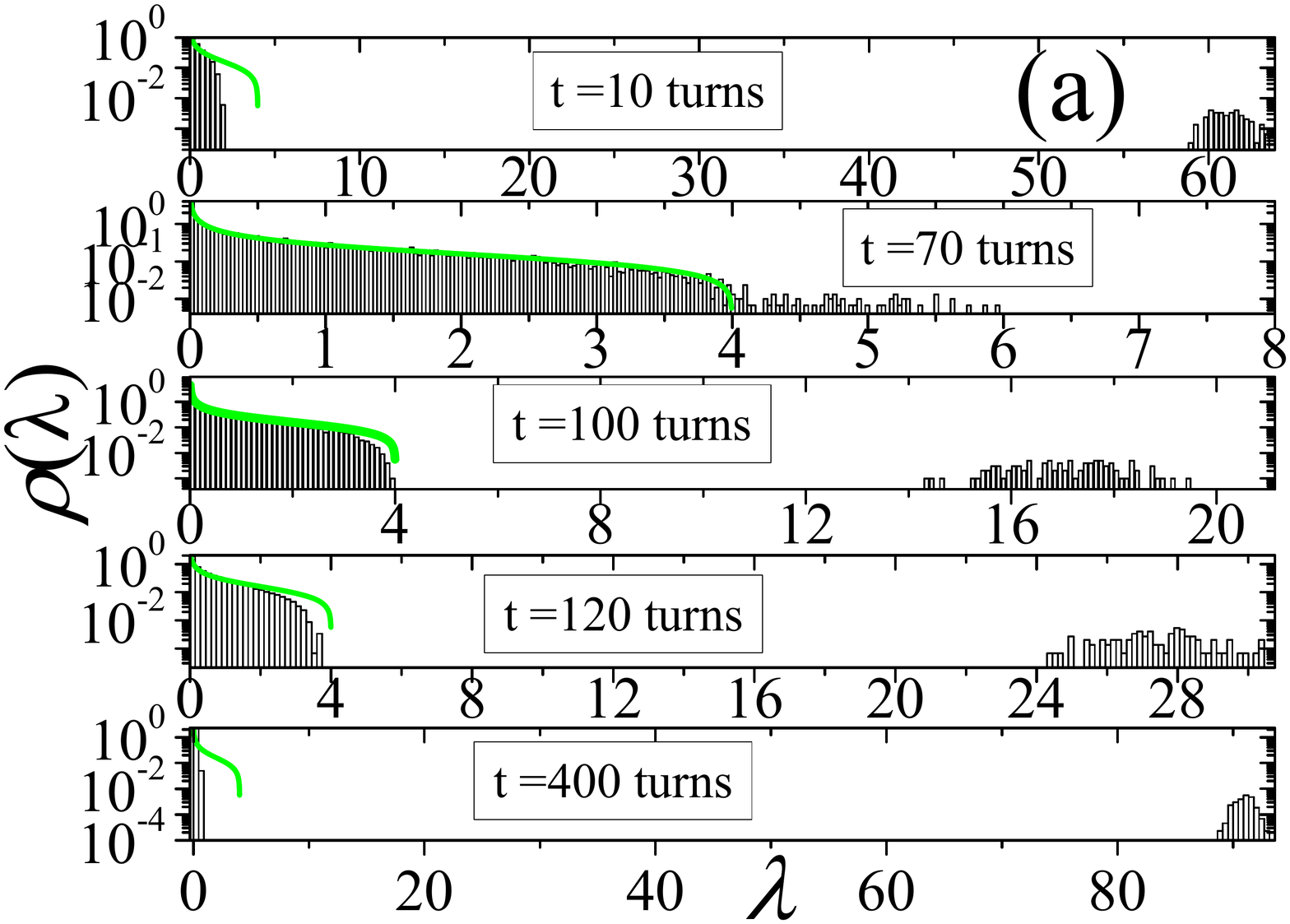} %
\includegraphics[width=1.0\columnwidth]{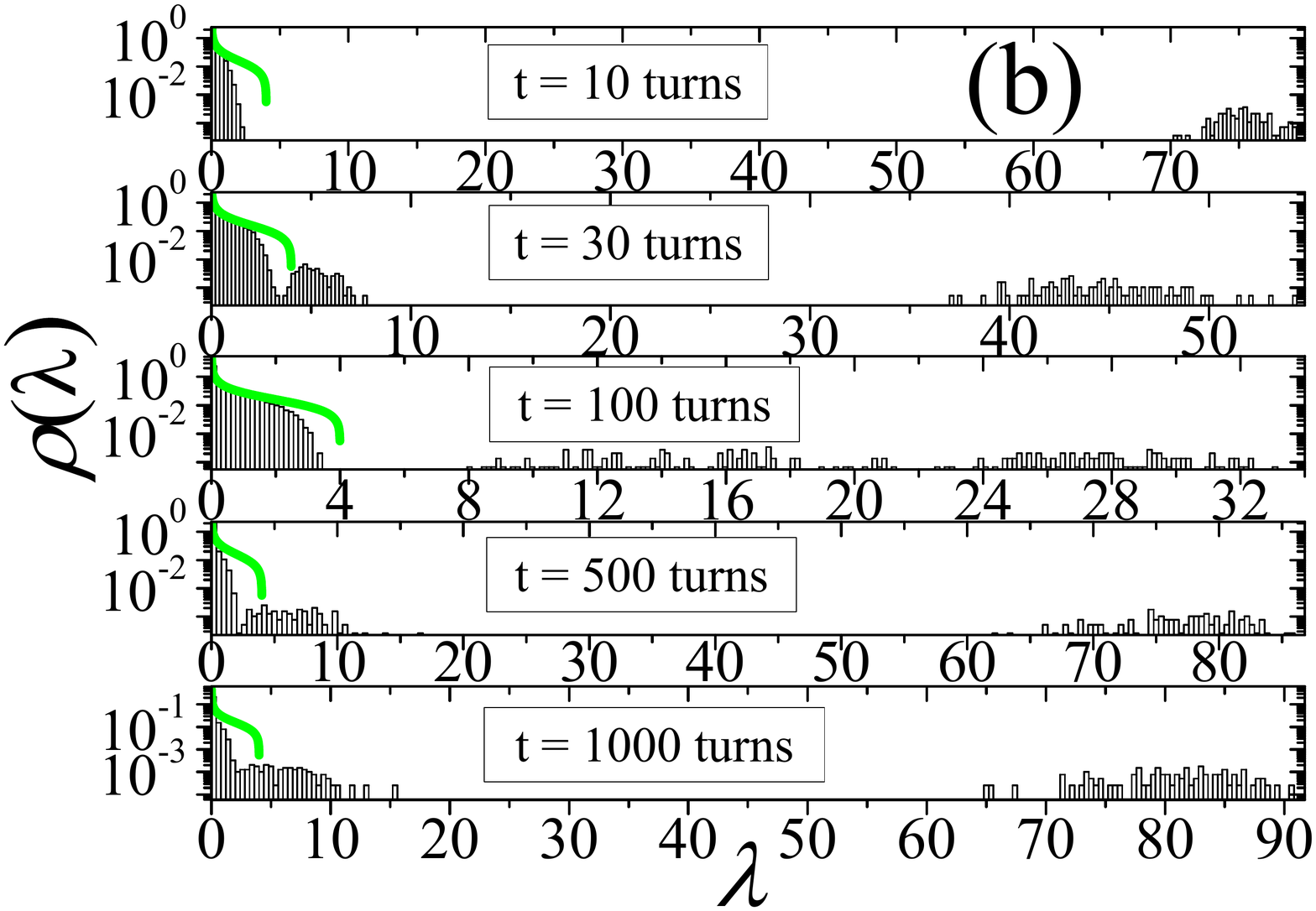}
\end{center}
\caption{Time evolution of the density of eigenvalues of the matrix $C$, for 
$N=100$, in two situations: a) No attractiveness evolution: $\protect\beta %
=1 $ b) With $\protect\beta $ evolving over time according to Eq. \protect
\ref{Eq:beta}. }
\label{Fig:Eigenvalues}
\end{figure}

Fig. \ref{Fig:Eigenvalues} (a) shows the time evolution of the density of
eigenvalues of the matrix $C$ obtained of the graph generated by random walk
with transitions where the attractiveness is fixed and equal to all nodes
during the evolution ($\beta =1$). We can observe that for $t\approx 70$
turns we have $\rho (\lambda )$ fitting the prescription of Eq. \ref%
{Eq:modified_semi_circle_law} (green continuous curve), which indicates that
we have all entries of adjacency matrix following approximately a symmetric
probability distribution, i.e., given two nodes there is an edge with
probability 1/2. After that the eigenvalues scape from bulk and at large
times all nodes are connected to all nodes (complete graph). In this
situation we have%
\begin{equation*}
\begin{array}{lll}
C_{\infty } & = & \frac{1}{N}\left( 
\begin{array}{cccc}
-1 & 1 & \cdots & 1 \\ 
1 & -1 & \ddots & \vdots \\ 
\vdots & \ddots & \ddots & 1 \\ 
1 & \cdots & 1 & -1%
\end{array}%
\right) ^{2} \\ 
&  &  \\ 
& = & \frac{1}{N}\left( 
\begin{array}{cccc}
N & N-2 & \cdots & N-2 \\ 
N-2 & N & \ddots & \vdots \\ 
\vdots & \ddots & \ddots & N-2 \\ 
N-2 & \cdots & N-2 & N%
\end{array}%
\right)%
\end{array}%
\end{equation*}%
that have only two distinct eigenvalues: $\lambda _{\min }=\frac{4}{N}$ and $%
\lambda _{\max }=\frac{(N-2)^{2}}{N}$ and for large times we have: $\rho
_{\infty }(\lambda )=c_{1}\delta (\lambda -\lambda _{\min })+c_{2}\delta
(\lambda -\lambda _{\max })$. For $N=100$, we have $\lambda _{\min }=0.04$
and $\lambda _{\max }\approx 96$. We can observe two peaks exactly in these
two values for $t=400$ turns.

Differently when the attractiveness evolves over time, plot (b) in Fig. \ref%
{Fig:Eigenvalues}, $\rho (\lambda )$ does not follow the law\ described by
Eq. \ref{Eq:modified_semi_circle_law} at any time since the evolution of
attractiveness leads to a situation where a lot of eigenvalues scape the
bulk. At large times we have not a complete graph since there are persistent
sites and the final distribution is not composed by two pronounced peaks as
expected in conditions of plot (a). More precisely we can understand that
for $\beta $ fixed, the graph can be mapped by random graph from connexion
probability $p=0$ until $p=1$ where different $p$ values corresponds to
different times of the evolution. Differently when $\beta $ evolves over the
time, this map is not obtained and a split of the main bulk of eigenvalues
can be observed differently from the case which $\beta $ is fixed. The
dynamics of eigenvalues scattering is also different and does not lead to
two peaks at large times since the network does not converge to a complete
graph. It is important to notice that the idea to characterize networks via
spectral analysis is not novel (see for example \cite{Albert2002}). The
scape of eigenvalues of the bulk for example were analysed in the context of
stock market analysis \cite{Stanley2000}. There, the authors use this
deviates from the expected law to characterize genuine correlations in the
stock market time series. Here we use to characterize the existence of
persistent nodes and consequently the deviation from the symmetric ($p=1/2$)
random graph behaviour.

\subsection{Crossover in $\protect\delta $}

In previous sections of this paper, we fixed $\delta =0$ and studied the
properties of the graphs generated by the spatial diffusion with evolution
of the attractiveness which leads to the existence persistent nodes. In
order to better understand the effects of $\delta $, we analyze two
important amounts, $p_{\infty }$ and the variance of attractiveness
distribution of the nodes, $var_{\infty }(\varphi )=\left\langle \varphi
^{2}\right\rangle -\left\langle \varphi \right\rangle ^{2}$ in the steady
state, for different systems size.

\begin{figure}[th]
\begin{center}
\includegraphics[width=1.0\columnwidth]{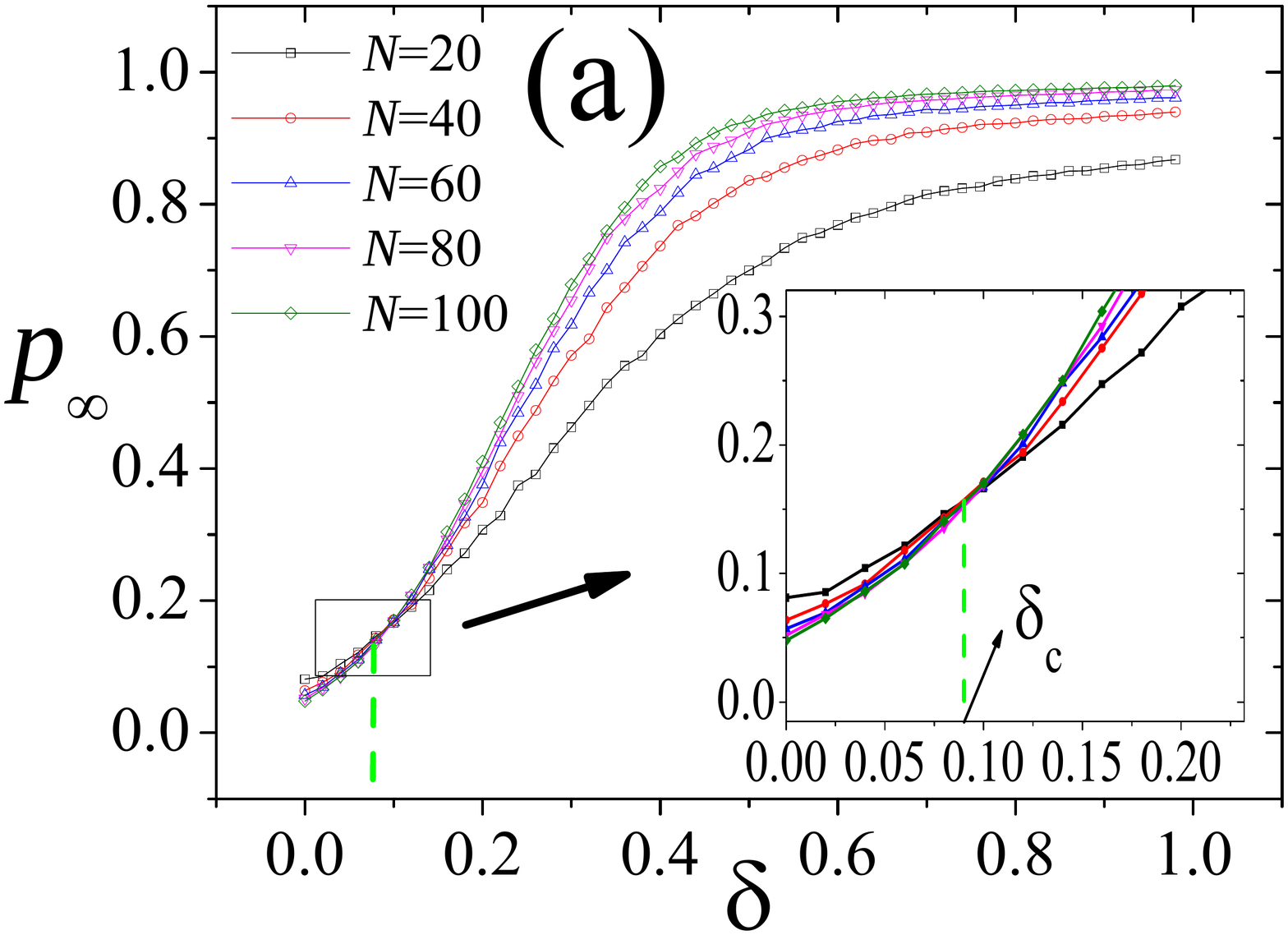} %
\includegraphics[width=1.0\columnwidth]{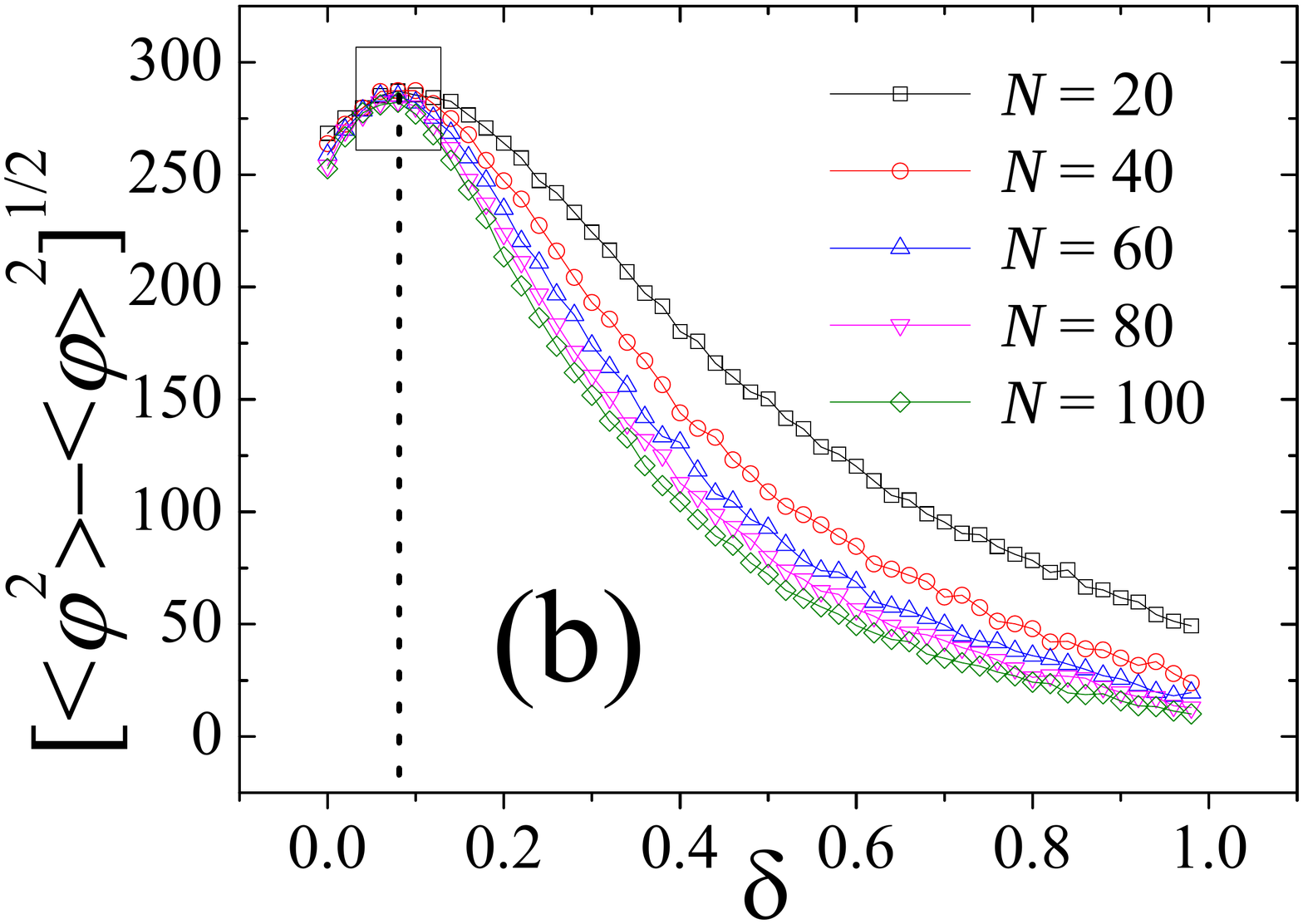}
\end{center}
\caption{(a) $p_{\infty }$ versus $\protect\delta $ for different systems
size. (b) variance of attractiveness as function of $\protect\delta $. Both
quantities were obtained in the steady state. }
\label{Fig:delta_effects}
\end{figure}

We can observe in Fig. \ref{Fig:delta_effects} the effects of $\delta $ on $%
p_{\infty }$ and $var(\varphi )$. First it is interesting to observe in plot
(a) that curves $p_{\infty }$ $\times $ $\delta $ for different system
sizes, have a crossover in $\delta \approx 0.09$. This crossover corresponds
exactly to the maximum in plot (b) for $var(\varphi )$ versus $\delta $.

\subsection{Comparing persistence for different transition probabilities}

An important question is to observe if the distance has some importance to
the existence of persistence nodes at large times as well as the other
transition probabilities. So we analyze different points:

\begin{enumerate}
\item Instead of $p_{ij}=\exp (-\frac{\varphi _{j}}{\varphi _{i}}d_{ij})$ we
consider $p_{ij}=\exp (-\frac{\varphi _{j}}{\varphi _{i}}\left\langle
d\right\rangle )\approx \exp (-0.519\frac{\varphi _{j}}{\varphi _{i}})$, in
order to observe if the effect of different distances has some importance on
the persistence;

\item We alternatively test non-Boltzmannian transition probabilities: $%
p_{ij}=\min \{1,\frac{\varphi _{j}}{\varphi _{i}}\}$, and $p_{ij}=\frac{%
\varphi _{j}}{\varphi _{i}+\varphi _{j}}$, which also follows the same idea:
how bigger $\varphi _{j}$ in relation to $\varphi _{i}$, bigger is $p_{ij}$.
However this functional dependence has no dependence on this distance
between nodes.
\end{enumerate}

\begin{table}[tbp] \centering%
\begin{tabular}{l|ll}
\hline\hline
\textbf{Transition probability} &  & \textbf{Formulae} \\ \hline\hline
I &  & $p_{ij}=\exp (-0.519\frac{\varphi _{j}}{\varphi _{i}})$ \\ 
&  &  \\ 
II &  & $p_{ij}=\exp (-\frac{\varphi _{j}}{\varphi _{i}}d_{ij})$ \\ 
&  &  \\ 
III &  & $p_{ij}=\min \{1,\frac{\varphi _{j}}{\varphi _{i}}\}$ \\ 
&  &  \\ 
IV &  & $p_{ij}=\frac{\varphi _{j}}{\varphi _{i}+\varphi _{j}}$ \\ 
&  &  \\ 
V &  & $p_{ij}=\exp (-d_{ij})$ \\ \hline\hline
\end{tabular}%
\caption{Transition probabilities}\label{Table:Transition_probabilities}%
\end{table}%

So we plot the time evolution of persistence considering different
transition probabilities.

\begin{figure}[th]
\begin{center}
\includegraphics[width=1.0%
\columnwidth]{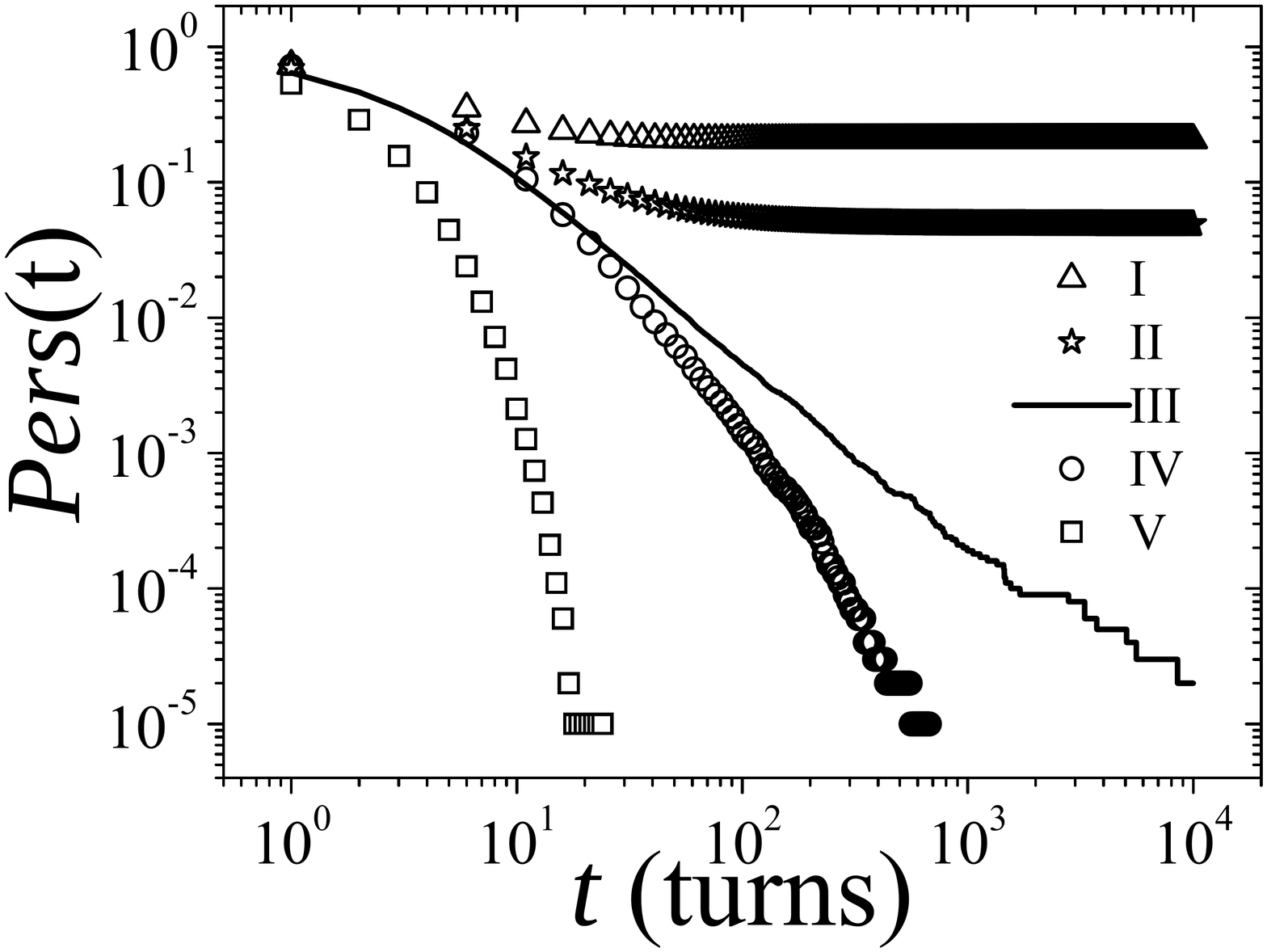}
\end{center}
\caption{$Pers(t)\times t$ for different transition probabilities (see Table 
\protect\ref{Table:Transition_probabilities}). We can observe for
exponential transition probabilities with $\protect\beta _{ij}=\protect%
\varphi _{j}/\protect\varphi _{i}$ that $Pers(t\rightarrow \infty
)=p_{\infty }$ (I and II). However we have that asymptotic value $p_{\infty
}^{(I)}$, which corresponds to the situation where $d_{ij}$ is changed by $%
\left\langle d\right\rangle \ $in the formula, is larger than $p_{\infty
}^{(II)}$. Finally we find power law decay of the persistence for a peculiar
non-Boltzmann transition probability (III). Exponential decays are observed
for $p_{ij}^{(IV)}$ and $p_{ij}^{(V)}$. The choice $p_{ij}^{(V)}$
corresponds to the case where the attractiveness does not change ($\protect%
\beta $ fixed equal to 1). The case IV, corresponds to other simple test of
non-exponential probability transition. This case, which also depends only
on attractiveness and not on the distances as the case III, leads to an
exponential decay and not a power law, although weaker than case V. }
\label{Fig:pers_diferentes_probabilidades}
\end{figure}

Fig. \ref{Fig:pers_diferentes_probabilidades} shows $Pers(t)\times t$ for
different transition probabilities (defined in Table \ref%
{Table:Transition_probabilities}) . We can observe for exponential
transition probabilities with $\beta _{ij}=\varphi _{j}/\varphi _{i}$ that $%
Pers(t\rightarrow \infty )=p_{\infty }$ (for both cases: I and II). However
we have $p_{\infty }^{(I)}$, which corresponds to the situation where $%
d_{ij} $ is changed by $\left\langle d\right\rangle \ $in the formula, is
larger than $p_{\infty }^{(II)}$, where the euclidean distances are indeed
considered. This test suggests that distance has importance for example on
the changing of the number of persistent sites in the steady state but not
on the decay behaviour that seems to be originated from two combined
aspects: the influence of attractiveness evolution in the transition
probability in an exponential Boltzmann formulae.

At this point an important question should be related to the existence of
power law decay as Eq. \ref{Eq.power_law_decay} for some $p_{ij}$? Thus, we
consider a non-Boltzmann probability transition: $p_{ij}^{(III)}=\min \{1,%
\frac{\varphi _{j}}{\varphi _{i}}\}$. In this case we can observe a power
law decay for the persistence, which suggests that some nodes should take a
very large time to be reached by the random walk according to this
prescription.

Exponential decays are observed for $p_{ij}^{(IV)}$ and $p_{ij}^{(V)}$. It
is important to see, that $p_{ij}^{(IV)}$ also corresponds to the previously
case where the attractiveness does not change ($\beta $ fixed equal to 1).
The case IV, corresponds to an alternative simple test to III of
non-Boltzmann probability transition. In this case as well as the case III,
only the attractiveness is important and the distances are not taken into
account. It is interesting to observe that we also obtain an exponential
decay for the persistence differently from III, but weaker than V where the
attractiveness is not important and only the distance is considered in the
Boltzmann rate.

\section{Summaries and Conclusions}

\label{Sec:Conclusions}

In this paper we explore the properties of an interesting preferential
attachment random walk over the points randomly distributed in a two
dimension unit square where transition probability depends on the distance
between nodes and also on attractiveness of the nodes which increases
according to the incidence of the random walks on these nodes.

We show that stationary persistent nodes (nodes that never were reached by
the random walk) appear when we modulate the transition probability with a
ratio $\beta $, which is given by attractiveness of arrival node divided by
the attractiveness of the departure node. When $\beta $ is fixed the
persistence exponentially decays to zero and no persistent node is observed.
This reason is deeply related to the fact that when $\beta $ is fixed the
network generated evolves passing to have characteristics of a symmetric
random graph up to reach a complete graph. Differently when $\beta $ evolves
considering the evolution of attractiveness of the nodes, the graph neither
passes through a symmetric random graph during the evolution nor reach a
complete graph as $t\rightarrow \infty $. This is corroborated by
distortions in the density of eigenvalues, degree distribution, and
attractiveness distribution.

We also find a crossover for the fraction of persistent sites for $%
t\rightarrow \infty $ between $p_{\infty }=0$ and $p_{\infty }=1$, for the
parameter $\delta $ that modulates the argument of the exponential that
describes the probability transition between two nodes prescribed by Eq. \ref%
{Eq:transition}. Our results suggest $\delta _{c}\approx 0.08$ for this
crossover which is corroborated with maximum of the variance of the
attractiveness of the nodes.

Finally we show that persistence can follow a power law decay for a specific
choice of non-Boltzmannian transition probability which does take into
account distances between the nodes and only the attractiveness of these
nodes. It is important to mention that another non-Boltzmann transition
probability used in this paper leads to exponential persistence exactly as
the Boltzmann one with $\beta $ fixed and taking into account the distances
between the nodes.

\textbf{Acknowledgments --} This research work was in part supported
financially by CNPq (National Council for Scientific and Technological
Development). We would like to thank Prof. S. D. Prado (IF-UFRGS) for kindly reading this manuscript and pointing out some interesting observations.

\end{document}